\documentstyle[epsfig,amssymb,onecolumn]{mn}

\newif\ifAMStwofonts
\title[Dynamical friction in dwarf galaxies]{An extensive study of dynamical friction in dwarf galaxies:
the role of stars, dark matter, halo profiles and MOND}
\author[S\'{a}nchez-Salcedo et al.]
{F.~J.~S\'anchez-Salcedo\thanks{E-mail:jsanchez@astroscu.unam.mx},
Jorge Reyes-Iturbide and X.~Hernandez\\
Instituto de Astronom\'{\i}a, UNAM, Ciudad Universitaria,
Apt.~Postal 70 264, C.P. 04510, Mexico City, Mexico}
\begin{document}

\date{Accepted xxxx Month xx. Received xxxx Month xx; in original form
2006 January 20}
\pagerange{\pageref{firstpage}--\pageref{lastpage}} \pubyear{2002}
\maketitle

\label{firstpage}

\begin{abstract}
We investigate the in-spiraling timescales of globular clusters in 
dwarf spheroidal (dSph) and dwarf elliptical (dE) galaxies, 
due to dynamical friction. 
We address the problem of these timescales having been variously 
estimated in the literature as much shorter than a Hubble time.
Using self-consistent two-component (dark matter and stars)
models, we explore mechanisms which may yield extended dynamical 
friction timescales in such systems in order to explain
why dwarf galaxies often show globular cluster systems. 
As a general rule, dark matter and stars both give a comparable contribution
to the dynamical drag. 
By exploring various possibilities for their gravitational make-up,
it is shown that these studies help constrain 
the parameters of the dark matter haloes in these galaxies, as well as to test
alternatives to dark matter. Under the assumption of
a dark halo having a central density core with a typical
King core radius somewhat larger than the observed stellar core radius,
dynamical friction 
timescales are naturally extended upwards of a Hubble time. 
Cuspy dark haloes yield timescales $\lesssim$ 4.5 Gyr, 
for any dark halo parameters in accordance with observations of 
stellar line-of-sight velocity dispersion in dwarf spheroidal galaxies.
We confirm, after a detailed formulation of the dynamical friction 
problem under the alternative 
hypothesis of MOND dynamics and in the lack of any dark matter, 
that due to the enhanced dynamical
drag of the stars, the dynamical friction timescales in MOND
would be extremely short.
Taking the well-measured structural parameters of the Fornax dSph and its
globular cluster system as a case study, we conclude that requiring 
dynamical friction timescales
comparable to the Hubble time strongly favours dark haloes
with a central core.

\end{abstract}

\begin{keywords}
galaxies: dwarf -- galaxies: individual (Fornax) -- galaxies:
kinematics and dynamics -- globular clusters: general -- gravitation
\end{keywords}

\section{Introduction}
The dynamical friction (DF) timescale for globular clusters (GCs) 
to sink to the centre
of most of dwarf spheroidals (dSph) and dwarf elliptical (dE) galaxies 
yields a fraction of the Hubble time, when dark halo parameters are 
taken straight from
measurements of the stellar distributions (e.g., Tremaine 1976; Hernandez
\& Gilmore 1998b; Oh, Lin \& Richer 2000; Lotz et al.~2001). 
Both the stellar bulk of the dwarf galaxy and the dark matter contribute
to the DF. In some dE, the DF
timescale with the dark matter alone is shorter than one Hubble time, 
in some dSph, the dynamical friction due to the stars alone would suffice 
to decay the GC orbits in a few Gyr, under the above assumption.
Tremaine (1976) first noticed that using the preliminary
values for the radius and mass of the second most luminous of the
$10$ dSph satellites of the Milky Way, Fornax, this time is $\sim 1$--$2$ Gyr,
very short as compared to absolute ages estimated for these clusters
($\sim 14.6\pm 1.0$ Gyr for clusters $1$--$3$ and $5$, $\sim 11.6$ Gyr
for cluster $4$, see Buonanno et al.~1998 and Mackey \& Gilmore 2003).
One would expect the formation
of bright nuclei from the merger of orbitally decayed GCs.
Paradoxically, Fornax contains five GCs,
an unusually high GC frequency for its
dynamical mass, and does not present a central nucleus. 
Although one of the clusters is observed located near the centre,
it presents a radial velocity comparable to the mean velocity of the
field stars (Dubath et al. 1992). 
Hence, this globular cluster has not been really dragged to the
Fornax centre through dynamical friction. 

Lotz et al.~(2001) found that the nuclei in $M_{V}>-14$ dE's are 
several magnitudes fainter than expected\footnote{Note that Fornax
has a magnitude $M_{V}=-12.5$.}.
Indeed, there are dE with faint nuclei 
and high globular cluster specific frequencies.
Therefore, either the DF timescale is being underestimated or,
in these galaxies, some mechanisms are working against DF
to prevent the GC system from collapsing into a 
bright nucleus (see Lotz et al.~2001 for a discussion of the
different possibilities).

In this paper we reconsider the problem of the orbital decay of globular
clusters in dwarf galaxies. In the case of central
cluster galaxies, a variety of processes may contribute to
altering the kinematics of GCs which are beyond the scope of
this paper. Using Fornax as a case study, we examine the
orbital evolution of GCs in both the conventional dark matter scenario and
under MOdified Newtonian Dynamics (MOND; Milgrom 1983). 
Although our work is aimed at exploring possible explanations for the
survival of GCs in dSph and dE,
our analysis will be focused mainly on the Fornax dSph 
because the recent detailed determinations of the structural
and dynamical parameters of both Fornax and its GC population
(i.e.~position, age, mass and radial velocity of the GCs are known),
allow us to test our assumptions.
Fornax is the prototypical dSph with a high GC frequency, has no
nucleus, and short DF timescale estimates. 

Since DF is a gravitational effect, our analysis is relevant for
any massive perturber, such as black holes, star clusters or any
kinematically cold massive substructure (e.g., Kleyna et al.~2004)
and not only for GCs.
Studies of the survival of GC systems and
density substructure can provide useful constraints
on the dark halo profile in dSph and dE (e.g., Hernandez \&
Gilmore 1998b), and could shed light on the debate of cuspy/cored haloes,
which is currently a test for the standard $\Lambda$CDM paradigm.
On the other hand,
since modified Newtonian dynamics (MOND; Milgrom 1983) can explain the
dynamics of spiral galaxies without any dark matter (e.g., Sanders 1996;
McGaugh \& de Blok 1998), and perhaps
the dynamics of dSph (Lokas 2001, 2002), one could ask whether MOND offers
a more natural solution to the problem of the rapid orbital
evolution of the globular
clusters in dwarf galaxies. Since MOND is solely determined by
the luminous material, the DF
timescale can be used as a test of modified dynamics. 

In \S 2 we give a general statement of the problem, reviewing
estimates of the DF timescales for the GCs in dwarf
galaxies, and discussing the processes that may be
working against the orbital decay of GC systems and their associated
difficulties. Section 3 presents a solution to the dynamical-friction problem
in a self-consistent two-component dynamical system, dark halo and stars,
giving analytical approximate solutions and full numerical integrations
for a range of possible dark halo profiles. In \S 4 we explore the resulting
in-spiraling timescales in the MOND scenario,
for various relevant limits of such theory.
We find that the problem of the sedimentation of GCs
has a simple solution in the dark matter
scenario, whereas it appears problematic in the MOND scenario.
Our conclusions are summarized in \S 5.

\section{Dynamical friction in dwarf galaxies. Statement of the problem}
\label{sec:statement}
In the standard Newtonian problem, the deceleration felt by a massive
perturber, e.g.~a globular cluster, of mass $M_{p}$ moving 
at velocity $\vec{V}$, 
produced by background particles of mass $m\ll M_{p}$ and
having an isotropic distribution function $f(v)$
is given by
\begin{equation}
\frac{d\vec{V}}{dt}=-16\pi^{2}\ln\Lambda G^{2} m M_{p}\frac{\vec{V}}{V^{3}}
\int_{0}^{V}f(v) v^{2} dv,
\label{eq:dvdt}
\end{equation}
where $\ln\Lambda$ is the Coulomb logarithm,  
with $\Lambda\sim b_{\rm max}/b_{\rm min}$ and $b_{\rm max}$ and
$b_{\rm min}$ a maximum and minimum impact parameters relevant to the problem.
For an extended perturber with internal velocity dispersion $\sigma_{p}$,
the minimum impact parameter is taken as
$b_{\rm min}\approx GM_{p}/\sigma_{p}^{2}$.
In the case of a GC moving on a circular orbit in 
centrifugal equilibrium within the core radius
$r_{0}\equiv (9\sigma^{2}/4\pi G\rho_{0})^{1/2}$ 
of a dwarf galaxy, where $\sigma$ is the one-dimensional velocity dispersion of
field particles and $\rho_{0}$ the central density,
the characteristic orbital decay timescale, derived in the local
approximation, is
\begin{equation}
t_{\rm df}=\frac{1}{\sqrt{3}\ln\Lambda}
\left(\frac{\sigma}{1\,{\rm km\,s^{-1}}}\right)
\left(\frac{r_{0}}{1\,{\rm kpc}}\right)^{2}
\left(\frac{M_{p}}{10^{5}\,{\rm M}_{\odot}}\right)^{-1}{\rm Gyr},
\label{eq:tremaine}
\end{equation}
(e.g., Tremaine 1976; Hernandez \& Gilmore 1998b; Oh et al.~2000).

In Fornax, the observed stellar velocity is $\sigma\approx 11$ km s$^{-1}$ 
and the luminous core radius $r_{0}=15'$, equivalent to almost $\sim 0.6$ kpc 
for an adopted distance of $138$ kpc (Mateo 1993; Walcher et al.~2003).
Adopting $b_{\rm max}=r_{0}$
and $b_{\rm min}=GM_{p}/4\sigma_{p}^{2}$ (e.g., White 1976), 
$\ln\Lambda\approx 3$--$4$ for
a typical Fornax GC of mass $2\times 10^{5}$ M$_{\odot}$
and central velocity dispersion $7$ km s$^{-1}$ (Dubath et al.~1992).
In order to give estimates, we will adopt a value $\ln\Lambda=3$.
Therefore, $t_{\rm df}\simeq 0.5$ Gyr for such a GC.
Since for eccentric orbits
with the same energy $t_{\rm df}$ is even shorter (e.g., Lacey \& Cole
1993; Colpi et al.~1999), it would appear that clusters are 
expected to sink toward the nucleus of the galaxy due to the dynamical friction
with the underlying stellar population (Tremaine 1976; Oh, Lin \& Richer 2000).
Contrary to this prediction, Fornax possesses five GCs
of $14.6$ Gyr of age (except cluster 4 which has an age of $11.6$ Gyr)
at $(1.60, 1.05, 0.43, 0.24, 1.43)$ kpc from the optical centre
and masses $(0.37, 1.82,
3.63, 1.32, 1.78)\times 10^{5}$ M$_{\odot}$, respectively (Mackey
\& Gilmore 2003), but no nucleus. 
The unusually high globular cluster frequency for its present mass suggests
that it was able to preserve its GC system intact without coagulation
of GCs in its centre. 

Tremaine (1976) proposed that once GCs reach the centre of
the galaxy they are ejected 
by the slingshot instability in a three-body system.
However, this expelling mechanism is only effective for point objects.
Decayed GCs should all merge together
forming a bright nucleus.

Oh, Lin \& Richer (2000) have attempted to identify
the tidal fields as a possible heating mechanism for GCs in dwarf
galaxies. In clusters of galaxies, the tidal interaction 
between the dE and the tidal field of the entire 
cluster of galaxies may be one possible stirring source.
In the case of Fornax, the tidal influence of the Milky Way may inject kinetic
energy into the dispersive motion of the GC population.
Indeed, Mayer et al.~(2001) have shown that the effects of the Galactic 
tidal field upon the local satellites can transform an extended 
low surface brightness dwarf into a dSph.

However, Oh et al.~(2000) show that the Galactic tidal effect by itself 
cannot counter the dynamical-friction drag within 
Fornax because the latter process operates on a shorter timescale.
Only if Fornax has experienced continuous and 
significant mass loss ($50$--$90\%$) would the clusters be stirred to
their actual distances. This scenario would imply that Fornax is undergoing
tidal disintegration and, consequently, a stream of tidally stripped material,
asymmetries in the shape of density profile and in its stellar population,
are expected. 
Since Fornax is at perigalactic, as dynamical studies strongly suggest, 
we should be observing the maximum effect of the tidal
interaction right now.
However, new wide field data reveals that Fornax has an almost perfect
relaxed spherical King profile, and no direct sign of interaction
with the tidal field of the Milky Way can be seen (Walcher et al.~2003).
The absence of any evidence of tidal stripping is in good agreement
with recent numerical work which suggests that tidal stripping
and shocking interior to $\sim 1$ kpc are both negligible 
for dSph galaxies with orbital pericentre $\gtrsim 30$ kpc and 
masses $\gtrsim 10^{8}$ M$_{\odot}$ (Read et al.~2006).

As remarked by Read et al.~(2006), the result that Local Group
dSph have not been significantly stripped interior to $1$ kpc
needs not invalidate the tidal transformation model presented
by Mayer et al.~(2001).
Indeed, once the galaxy has been
transformed into a dSph, the Galactic tidal field ceases to 
significantly affect the internal dynamics
of the dwarf. These would hence result in marginal internal tidal heating, 
as suggested by the relaxed kinematics and spherical present morphology 
of Fornax. 

To summarize this point, we can say that although Galactic 
tidal effects upon Fornax can not be completely ruled out, 
particularly over its poorly known initial stages, tidal heating 
appears insufficient to account for the survival of the observed GC system.

There are other potential sources of external dynamical heating of the GC
population, 
such as the tidal interaction with a hypothetical companion galaxy 
that finally merged with the dwarf.
Interestingly, there is some evidence of
a merger of a small ($10^{5}$--$10^6$ M$_{\odot}$) companion that occurred 
$\sim 2$ Gyr ago (Coleman et al.~2004). However, a system of that mass 
would be unable to tidally stir the GC population to the required degree.
Another possible mechanism for injecting kinetic energy into the GCs
is by scattering of the GCs by massive black holes (Oh 
et al.~2000). However, the gravitational encounters would disrupt the clusters
(e.g., Lotz et al.~2001). Moreover, this scenario predicts some features that
are not observed (Oh et al.~2000).
Since the drag depends on the mass of the
GCs, one could suggest that the GCs formed only
recently by successive mergers of less massive GCs.
Current observations indicate that
the clusters in Fornax are not likely merged objects.

A similar analysis for the GCs in 
NGC 185 (Tremaine 1976) and dE (Lotz et al.~2001) leads
to the same conclusion: if the GCs formed with the same radial
profile as the stellar component, the entire GC system would collapse
into a single bright nucleus. 
It is indeed unlikely that all the GCs have been captured recently by the
parent dE or dSph or that they constitute a young population, 
so that the dynamical-friction drag had no time to
spiral them into the centre.
For a more detailed discussion of the difficulties of the above scenarios,
we refer the reader to Oh et al.~(2000) and Lotz et al.~(2001).  
In conclusion, the extended distributions of GCs in Fornax and
dE's need an explanation. 

Notice that in going from Eq.~(\ref{eq:dvdt}) to Eq.~(\ref{eq:tremaine}) 
it has been assumed that the circular velocity of the
object experiencing the dynamical friction is in equilibrium 
with the potential produced by the 
same distribution of background field particles $f(v)$ 
which cause the deceleration.
Hence, the core radius of Eq.~(2) should be that of the density 
distribution responsible for
the observed $\sigma$. The estimates referred to at the start of this 
section however, consider
only one dynamical component, inconsistent in the absence of a self-consistent 
gravitational model, as the large $M/L$ values of
these systems signal. As it will become clear in the following sections, a consideration of the dynamical
friction problem in a two-component self-consistent model allows to 
reconcile the observed dwarf galactic
parameters with the expectation of long DF timescales for 
GCs in these systems.

\section{Examining the two-component dynamical friction}
Equation (\ref{eq:tremaine}) provides the DF timescale
of a body orbiting within the core of a dSph due to
the combined action of field stars and dark matter but
only under the assumption that mass-follows-light. 
We know, however, that dark matter and
stars may have different distribution functions; on no galactic
scales does mass follow light. In the case of Fornax, it is clear
that a single component King model is
invalid (e.g., Walker et al.~2005). Hence the use
of Eq.~(\ref{eq:tremaine}) is not justified.
As suggested by Hernandez \& Gilmore (1998b), an increase in the
dark matter core will increase the velocity dispersion of the dark matter
and decrease the DF with these particles (but not with the stars).
It would be desirable to explore if there exists a reasonable range of halo 
parameters, consistent with the observed stellar projected density
and kinematics, for which the dynamical friction is significantly
suppressed.

As the DF force depends on the distribution function, we would need
self-consistent distribution functions for the stars and dark 
matter particles moving in the same underlying matter potential.
The usual way to proceed is to suppose a form for the distribution
functions and constrain the parameters of the model  
from the measured light distribution and the velocity dispersion profile
of the stellar component. 
Because the King model is known to provide a good fit
to the present light distribution of the inner regions of certain dSph
and galactic dark haloes (Hernandez \& Gilmore 1998a), 
a two-component King model could be a natural choice
(e.g., Pryor \& Kormendy 1990).

If $E=-v^{2}/2-\phi$ is the total (positive) energy per unit of 
mass and $v$ the particle velocity, each component $j$ 
is assumed to have the energy distribution function:
\begin{equation}
f_{j}(E)\propto e^{-\mu_{j}\left(E-v_{0}^{2}W_{0}\right)/v_{0}^2}-1,
\label{eq:distributions}
\end{equation}
where $v_{0}$ sets the velocity scale and $W_{0}$ is the central depth of the 
potential in units of $v_{0}^{2}$. For luminous (subscript ${\star}$) and
dark (subscript ${\rm dm}$) components, 
the model has five parameters: $r_{0}$ (the scale radius), $v_{0}$, 
$W_{0}$, ${\mathcal{R}}$ (defined as the ratio of central densities
$\rho_{0,dm}/\rho_{0,\star}$) and $\mu_{dm}/\mu_{\star}$.
The $\mu_{j}$ have a physical interpretation
when multicomponent models are applied to stellar mass classes. For
our purposes, $\mu_{dm}/\mu_{\star}$ determines the ratio between
the King core radii of dark matter and stars, $\beta$.
More specifically, $r_{c,j}^{2}=r_{0}^{2}/\mu_{j}$ and thus
$\beta\equiv \hat{r}_{\rm dm}/\hat{r}_{\star}=(\mu_{\star}/\mu_{dm})^{1/2}$
(Gunn \& Griffin 1979).

The distribution function has the form $f=\sum_{j} f_{j}$, implying that
the dynamical-friction force is the linear sum of the contributions
of stars and dark particles.
Since the problem of the decay of GCs concerns those that are not
far from the centre, we will focus on the behaviour of the volume
density and velocity dispersion of stars and dark particles
only within a distance $\sim 2\hat{r}_{\star}$ from the 
galaxy centre. This greatly simplifies the discussion and physical 
interpretation.

Under Eq.~(\ref{eq:distributions}) it is easy to see that, 
within $\sim 2\hat{r}_{\ast}$, the velocity dispersion profiles of  
the stellar and dark matter components are flat as long as
$W_{0}>6$ and $\beta>1$. This can be seen from very basic grounds.
If $\beta=1$, stars and dark 
particles are isothermal, both of which have the same velocity dispersion,
 within $2\hat{r}_{\star}$, for $W_{0}>6$
(see Fig.~4-11 of Binney \& Tremaine 1987). If $\beta>1$, 
the velocity dispersion profiles start to decline at larger radial distances
(e.g., Pryor \& Kormendy 1990; Walker et al.~2005).
The reason is that $e^{-\mu_{j}(E-v_{0}^{2}W_{0})/v_{0}^{2}}$ 
is large compared to unity along a more extended radius range.
In fact, as commonly argued, the flat behaviour of the stellar velocity 
profile in many dSph galaxies arise naturally if the stars orbit inside a
dark matter halo with a King core radius larger than that of the visible
component ($\beta>1$). 
If, in addition, each component satisfies $\mu_{j}W_{0}>6$, 
the central velocity dispersion of 
each component $j$ is simply related to $v_{0}$ according to
$\sigma_{j}^{2}=\mu_{j}^{-1}v_{0}^{2}$ and
\begin{equation}
\sigma_{j}^{2}=\frac{4\pi G}{9}\rho_{0} r_{j}^{2},
\label{eq:coreradius}
\end{equation}
where $\rho_{0}=\rho_{0,{\rm dm}}+\rho_{0,\star}$
(Gunn \& Griffin 1979). The best fit dynamical two-component King models
in Fornax suggest  $7<W_{0}<9$,
with still larger values not being ruled out (Walker et al.~2005). 
Models with $W_{0}<7$ are excluded. Hence, the condition 
$\mu_{\star}W_{0}>6$ is fulfilled, bearing in mind that the more
concentrated component (the stars in our case) has $\mu_{j}>1$.
\footnote{We may exploit the fact that the
velocity dispersions $\sigma_{\rm dm}$ and $\sigma_{\star}$ are
constant within $r<2\hat{r}_{\star}$,
to find a relation between the dark matter and the stellar density profile
$\rho_{\rm dm}={\mathcal{R}}(\rho_{\star}/\rho_{0,\star})^{\lambda}$, where
$\lambda=\sigma_{\star}^{2}/\sigma_{\rm dm}^{2}\simeq 
\hat{r}_{\star}^{2}/\hat{r}_{\rm dm}^{2}$.}

Unfortunately, this family of models presents serious limitations 
because the assumption that the luminous and dark components are
dynamically coupled does not permit to have large values of $\beta$. 
In a recent careful analysis of stellar velocity dispersions in Fornax
using two-component King models, Walker et al.~(2005) 
found that the models able to reproduce the data span a narrow
range in size of the dark halo, $2<\beta<3$.
As discussed in detail by Walker et al.~(2005), the restriction
$\beta<3$ is an artifact of the two-component King model, since
for large $\beta$ it produces excessively steep luminous matter profiles 
(Pryor \& Kormendy 1989).

Since our goal is to obtain how the DF timescale depends
on $\beta$, we will adopt a different approach. We will assume
a certain form for the density profile of the stars and dark matter 
and derive the
velocity dispersion through a Jeans equation:
\begin{equation}
\sigma_{j}^{2}(r)=\frac{\rho_{0,j}\sigma_{j}^{2}(0)}{\rho_{j}(r)}
-\frac{1}{\rho_{j}(r)}\int_{0}^{r}
\rho_{j}(r') \frac{G(M_{\rm dm}(r')+M_{\star}(r'))}{r'^{2}} dr',
\label{eq:dispersion}
\end{equation}
where $M_{\star}(r)$ and $M_{\rm dm}(r)$ are the mass of stars and
dark matter within radius $r$, respectively.
Because the King model is known to provide a good fit
to the present light distribution of the inner regions of certain dSph, 
globular clusters and large galactic dark haloes (Hernandez \& Gilmore 1998a)
and to keep matters simple, we will assume that the distributions
of both the luminous component and the dark matter
follow a King profile, being the dark matter halo
more extended than the population of stars, so that
$\hat{r}_{\rm dm}$ will be larger than $\hat{r}_{\star}$.
With the present quality of the data, we found that
values $2<\beta < 5$ are consistent with the
line-of-sight velocity dispersion profile measured in Fornax.
Other profiles for the dark halo and their implications
will be explored in \S \ref{sec:profiles}.

For obvious reasons,
there have been many attempts to find an analytical fit to the luminosity
profile of GCs and dSph galaxies:
the modified Hubble law (1930), the empirical King profile (King 1962)
or the Plummer model (Lake 1990) are some examples.
In this section we
present two analytical approximations which yield good approximate fits to
King density profiles. The first is:
\begin{equation}
\rho(x)=\frac{\pi \rho_{0}}{4}\left(\frac{{\rm erf} (x)}{x} \right)^{2},
\label{eq:xavierapprox}
\end{equation}
where $x\equiv 3r/2\hat{r}$, $\rho_{0}$ the central density
and ${\rm erf}(x)$ is the error function, and the second:

\begin{equation}
\rho(x)=\frac{\pi \rho_{0}}{4 F}\left(\frac{{\rm erf} (x)}{x} \right)^{2},
\label{eq:xavierapprox2}
\end{equation}
where $F$ is a correction factor equal to 1.0 for $r/ \hat{r}<2.0$ and equal to
$2-(2\hat{r}/r)$ for  $r/ \hat{r}>2.0$. 

Figure 1 gives the logarithm
of the ratio of the density profiles of Eq.~(\ref{eq:xavierapprox}) and
Eq.~(\ref{eq:xavierapprox2}) to the exact King profiles, for the
eight values of the shape parameters 2, 4, 6, 8, 10, 12, 14 and 16, 
left to right. 
It can be seen that for all shape parameters larger than or 
equal to 4, Eq.~(\ref{eq:xavierapprox})
is an extremely accurate fit to the King profile within one core radius.
We see also that Eq.~(\ref{eq:xavierapprox}) is an excellent approximation 
to within a factor of 1.06 (log = 0.025) of the exact value, 
internal to two core radii, 
for all shape parameters larger than or equal to 8. This 
accurately takes into account the drop in density observed within 
the core radius of a King sphere, 
of a factor close to 0.5. It should be noted that this drop is much 
more substantial than the corresponding one in the velocity dispersion, 
which remains flat to within 
a few percent inside the core region  (Binney \& Tremaine 1987). 

For shape parameters between 12 and 14,
Eq.~(\ref{eq:xavierapprox2}) lies within a factor
of 1.04 of the exact King profile within 25 core radii. The two 
equations above, together with Figure 1, are included
as they might serve as accurate approximations to full King profiles 
in related work and are useful to derive analytical estimates. 
Our problem is restricted to the core region of the dark matter
distribution, so that in the following
only  Eq.~(\ref{eq:xavierapprox}) will be used.

\begin{figure*}
  \includegraphics[width=0.8\textwidth]{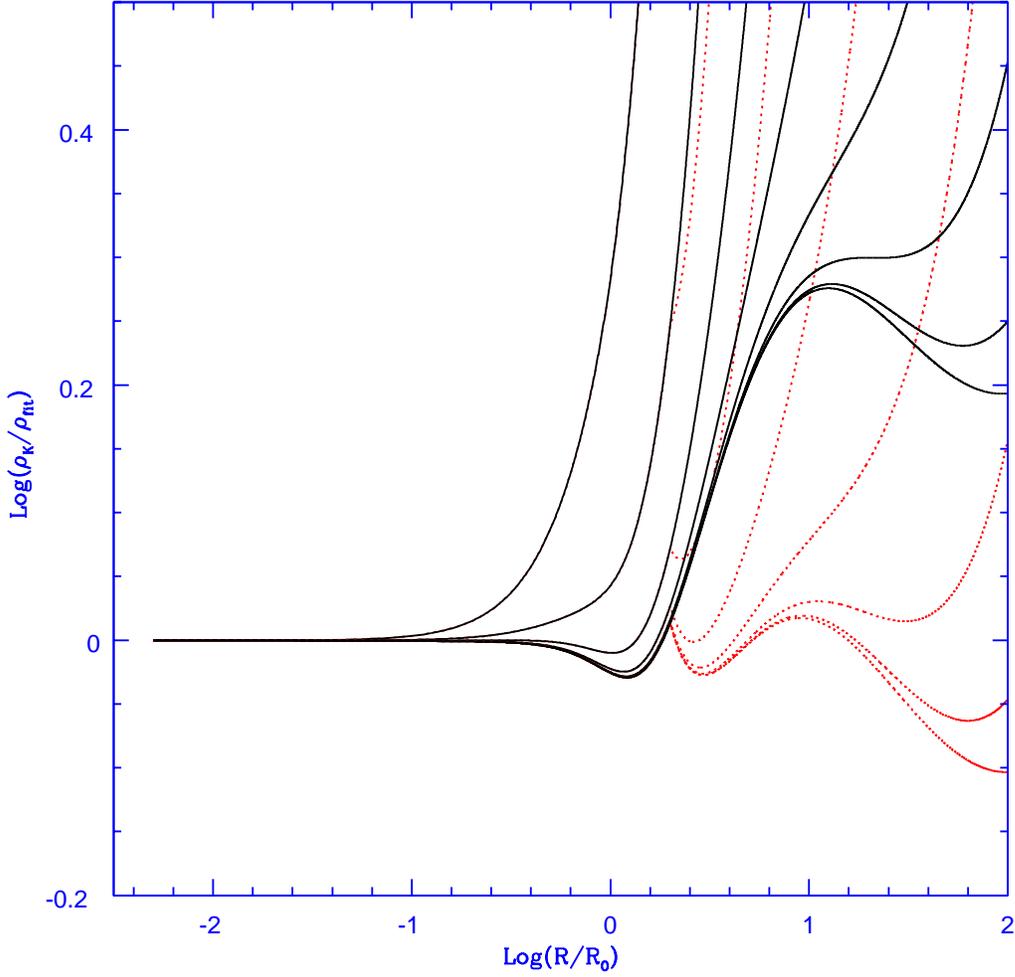}
  \caption{Logarithms of the ratios of the analytical
density profiles as given in Eqs (\ref{eq:xavierapprox}) and 
(\ref{eq:xavierapprox2}) to 
exact King profiles (solid and dotted lines, respectively)
with shape parameters of 2, 4, 6, 8, 10, 12, 14 and 16
(from left to right).}
  \label{fig:evolution}
\end{figure*}

\subsection{Dynamical friction in a two-component system}
\label{sec:basicequations}

Under the approximation that the GCs orbit at the circular speed
$v_{c}(r)$ as they spiral to the centre, the stellar component
contributes to the DF force with:
\begin{equation}
F_{\star}(r)=-\frac{4\pi \ln \Lambda G^{2} \rho_{\star}(r) M_{p}^{2}}
{v_{c}^{2}}\left[{\rm erf}\left(\frac{v_{c}}{\sqrt{2}\sigma_{\ast}}
\right)-\sqrt{\frac{2}{\pi}}\frac{v_{c}}{\sigma_{\ast}}\exp\left(
-\frac{v_{c}^{2}}{2\sigma_{\ast}^{2}}\right)\right],
\label{eq:dfstars}
\end{equation}
where $\sigma_{\star}(r)$ can be obtained from
Eq.~(\ref{eq:dispersion}) and the circular velocity is given by
\begin{equation}
v_{c}(r)=\sqrt{\frac{G(M_{\ast}(r)+M_{\rm dm}(r))}{r}}.
\end{equation}
An analogous form for the contribution of
the background of dark matter particles, $F_{\rm dm}(r)$,
can be derived by replacing $\rho_{\star}\leftrightarrow \rho_{\rm dm}$
and $\sigma_{\star}\leftrightarrow \sigma_{\rm dm}$.
Notice that in this approach there is no guarantee that the distribution 
function is strictly Maxwellian. However, the errors made in this
approximation are likely small.

As said before, we will use the profile
given by Eq.~(\ref{eq:xavierapprox}) for the underlying
stellar population, with central density $\rho_{0,\star}$, as well
as for the dark matter, with central density $\rho_{0,{\rm dm}}$.
Taking this analytical density profile, which is as accurate as a 
factor of $1.06$ in the interval $0<r<2\hat{r}$, provided that
the shape parameter $>8$, 
ensures that we incur in errors much smaller than the measured 
uncertainties in, for example, the stellar core radius or the 
inferred central matter density for Fornax.

For simplicity and to isolate the relevant physics, we will take
the Coulomb logarithm constant with a value $\ln\Lambda=3$,
regardless of whether the interaction occurs with dark particles or stars.
This is indeed an approximation since $\Lambda$ is expected to
be a linear function of the distance of the GC to the centre
of the galaxy, $\Lambda\sim r/b_{\rm min}$
(e.g., Hashimoto et al.~2003; Just \& Pe\~{n}arrubia 2005).
Therefore, our calculations overestimate the drag at small radii.
Note, however, that the absolute value of $\Lambda$ near the
stellar core radius
may vary in the range $3$--$4$ depending on the half-mass
radius of the GCs (see \S \ref{sec:statement}).

Once we know the friction force
we can solve numerically the orbital evolution of a GC.
Before presenting the results, we will give an estimate of the
characteristic DF timescale for a GC
within the core.

\subsection{Characteristic two-component orbit-decay timescale}
It is convenient to gain additional
insight by examining the drag felt by a GC bound to orbit within the stellar
core radius, where $v_{c}^{2}\approx
4\pi G(\rho_{0,\ast}+\rho_{0,{\rm dm}})r^{2}/3$.
Substituting into Eq.~(\ref{eq:dfstars}) and using 
$\sigma_{\star}^{2}= 4\pi G\rho_{0,\star}(1+{\mathcal{R}})
\hat{r}_{\star}^{2}/9$ as derived in the exact two-component
King model (see Eq.~\ref{eq:coreradius}),
the DF force by the stars can be written as:
\begin{equation}
F_{\star}(r)=-\frac{3 \ln \Lambda G M_{p}^{2}}
{(1+{\mathcal{R}})r^{2}}
\left[{\rm erf}\left(\sqrt{\frac{3}{2}}\frac{r}{\hat{r}_{\ast}}
\right)-\sqrt{\frac{6}{\pi}}\frac{r}{\hat{r}_{\ast}}\exp\left(
-\frac{3r^{2}}{2\hat{r}_{\ast}^{2}}\right)\right].
\label{eq:corefrictionst}
\end{equation}
From Eq.~(\ref{eq:corefrictionst}) we see that the stellar contribution
to the friction force is reduced by a factor $1+{\mathcal{R}}$, as
compared to the force if all the matter were concentrated in
one single component following the light distribution.
In the same approximation but for the dark matter component, we obtain:
\begin{equation}
F_{\rm dm}(r)\simeq -\frac{3 \ln \Lambda G M_{p}^{2}}
{(1+{\mathcal{R}}^{-1})r^{2}}
\left[{\rm erf}\left(\sqrt{\frac{3}{2}}\frac{r}{\hat{r}_{\rm dm}}
\right)-\sqrt{\frac{6}{\pi}}\frac{r}{\hat{r}_{\rm dm}}\exp\left(
-\frac{3r^{2}}{2\hat{r}_{\rm dm}^{2}}\right)\right].
\label{eq:corefrictiondm}
\end{equation}
For analytical purposes, we have used here the simplifying assumption that
in dark matter dominated systems, $\sigma_{\rm dm}^{2}\simeq
4\pi G\rho_{0,{\rm dm}}\hat{r}_{\rm dm}^{2}/9$, which is certainly
a lower limit.
After a Taylor expansion of Equation (\ref{eq:corefrictiondm})
for $r\ll \hat{r}_{\rm dm}$, we find
\begin{equation}
F_{\rm dm}(r)=-3\sqrt{\frac{6}{\pi}}\frac{\ln\Lambda GM_{p}^{2}}{1+{\mathcal{R}}^{-1}}
\frac{r}{\hat{r}^{3}_{\rm dm}}.
\end{equation}
This shows that
for relatively large values of $\hat{r}_{\rm dm}$,
corresponding to high $\sigma_{\rm dm}$,
the  friction force by the dark matter component may be suppressed
considerably, leaving a door open for the survival of the orbits
of GCs. The reduction of $F_{\rm dm}$ in extended haloes
was already highlighted by Hernandez \&
Gilmore (1998b) and Lotz et al.~(2001).

Within the luminous core of the dwarf galaxy, the total friction
force is:
\begin{equation}
F(r)=F_{\star}(r)+F_{\rm dm}(r)\simeq
-3\sqrt{\frac{6}{\pi}}\frac{\ln\Lambda GM_{p}^{2}}{1+{\mathcal{R}}}
\frac{r}{\hat{r}^{3}_{\star}}\left[1+\frac{{\mathcal{R}}}{\beta^{3}}\right].
\label{eq:TwoCompDF}
\end{equation}
As a measure of the rate of the orbit decay, we define $\tau_{5th}$
as the time for the GC orbit to decay to $1/5$th of its initial
radius.
The calculation of the characteristic timescale
$\tau_{5th}$ is similar to that set out in Binney \& Tremaine (1987).
As such, we only give the final result:
\begin{equation}
\tau_{5th}\simeq 0.77 \left[\frac{1+{\mathcal{R}}}
{1+{\mathcal{R}}/\beta^{3}}\right]\frac{v_{c}(\hat{r}_{\star})
\hat{r}_{\star}^{2}} {GM_{p}\ln\Lambda}
\simeq 1.33 \left[\frac{1+{\mathcal{R}}}
{1+{\mathcal{R}}/\beta^{3}}\right]\frac{\sigma_{\ast} \hat{r}_{\star}^{2}}
{GM_{p}\ln\Lambda}.
\label{eq:newtonian}
\end{equation}
The previous formula constitutes the generalization 
of Eq.~(\ref{eq:tremaine}) for the
two-component case, provided that the object resides well inside the
stellar core radius.
For values ${\mathcal{R}}\gtrsim 10$ and
$\beta\gtrsim 2$, the dynamical friction timescale can be significantly
enhanced by a factor $\gtrsim 5$. In particular, for ${\mathcal{R}}=10$,
$\beta=2$ and the above mentioned reference values
of $\sigma_{\star}=11$ km s$^{-1}$, $\hat{r}_{\star}=0.6$ kpc and
$M_{p}=2\times 10^{5}$ M$_{\odot}$ we get $\tau_{5th}\simeq 10$ Gyr.
We see that without violating stellar velocity dispersion constraints, 
dynamical friction timescales can be significantly extended by 
assuming cored dark matter haloes. By varying the King core radius
of the dark halo in the range $2<\beta<5$, we are changing the dark matter 
particle distribution function, increasing their assumed velocity 
dispersion and hence suppressing their
dynamical friction contribution, but the effect on the dark matter 
bulk gravitational force contribution within the halo core, 
responsible for the observed stellar kinematics, is small enough
to be entirely compatible with present quality of the data.
In the following we present the exact evolution of the
orbit for a GC in a dwarf spheroidal galaxy with a mass
and size similar to those of Fornax.

Returning to the opening discussion of \S 2, dynamical friction estimates
based on Eq.~(\ref{eq:tremaine}), taking the measured stellar core radii 
$\sim \hat{r}_{\ast}$ and velocity 
dispersions $\sigma_{\ast}$, implicitly assume that mass follows light,
and thus the friction drag with the dark matter component is included.
Therefore, even though Eq.~(\ref{eq:tremaine}) can be written in
terms of the stellar parameters only, it does not follow that the
drag caused by the stellar wake alone yields 
excessively short DF timescales. 

\subsection{Numerical results}
\label{sec:numerical}
As shown in the previous section, we expect that
for values of ${\mathcal{R}}$ and $\beta$ large enough, the problem
of the coalescence of GCs in the central region of the host
galaxy should be alleviated. 
Following the description given in \S \ref{sec:basicequations}, the evolution 
of the orbit of a GC was derived for a plausible range of parameters. 
We present the orbital decay for different values of
${\mathcal{R}}$, $\beta$ and the circular velocity of the total 
mass distribution at $\hat{r}_{\ast}$, denoted by $\hat{v}_{c}$,
keeping the stellar core radius fixed at the $0.6$ kpc measured for Fornax. 
In fact, the set $({\mathcal{R}},\beta,\hat{v}_{c},\hat{r}_{\ast}=0.6$ kpc) 
determines the mass model completely.
The velocity dispersion and $\hat{v}_{c}$ are related through
$\hat{v}_{c}\sim \sqrt{3}\sigma_{\ast}$. Thus,
observations of $\sigma_{\ast}$ provide an estimate of $\hat{v}_{c}$,
implying $\hat{v}_{c}\sim 20$ km s$^{-1}$ for the observed value
$\sigma_{\ast}\approx 11$ km s$^{-1}$ in Fornax.
The halo parameters for our calculations were chosen to span the
observational estimates of the mass of Fornax.
Lokas (2002) derived a $M/L=15$--$25$ inside a galactocentric radius
of $2$ kpc by modeling the moments of the line-of-sight velocity
distributions. 
Using projected positions and radial velocities of stars in Fornax,
Wang et al.~(2005) estimated a $M/L=7$--$22$ within a radius of $1.5$ kpc, 
and Walker et al.~(2005) suggest a cumulative $M/L=7$--$10$ within
the core radius.
Hence, ${\mathcal{R}}\sim 7$--$20$ is consistent with observations. 
These authors also show that 
Fornax contains an extended dark halo; we will explore values for 
$\hat{r}_{\rm dm}$ in a reasonable range of $\sim 1$--$2$ kpc,
which implies $\beta=1.6$--$3.2$.  

The evolution of the orbital radius of a GC of mass 
$2\times 10^{5}$ M$_{\odot}$ moving initially on a circular orbit, for
different mass models, is shown in Fig.~\ref{fig:evolution}.
Given the linear nature of Newtonian gravity, as seen in the preceding 
sections, the dynamical friction force will be the sum of that due to stars 
and dark matter, c.f.~Eq.~(\ref{eq:TwoCompDF}).
Still, these two components are coupled through the self-consistent galactic 
mass model, as the stellar velocity dispersion is largely determined 
by the dark matter component.
Considering the large overall mass to light ratios of dSph systems, 
one might naively think that dynamical friction is determined completely 
by the $F_{\rm dm}$ term in Eq.~(\ref{eq:TwoCompDF}),
and that the effect of stars can be ignored. However, the details of 
local density and different velocity 
dispersion of these two components, and the 
local rather than global nature of dynamical friction 
make things more complicated.
In order to illustrate this issue, we have
separated both contributions in panel (a) of Fig.~\ref{fig:evolution}.
The curve labeled with ``stars'' represents the radial evolution
if the dark halo acts as an unresponsive external potential so that
it does not produce any dynamical friction, i.e.~$F_{\rm dm}=0$.
The curve ``dark matter'' was calculated with $F_{\star}=0$. 
In both cases ${\mathcal{R}}=7$, $\hat{v}_{c}=20$ km s$^{-1}$ and $\beta=2.5$ 
(or, equivalently, a King core radius $\hat{r}_{\rm dm}=1.5$ kpc).
It is interesting to note that even though the density of
dark matter is a factor $\sim 7$ larger than the baryonic density,
both drag forces are comparable for a GC traveling
within the stellar core; both components must be included 
if an accurate description of the problem is to be obtained.

The evolution when $F_{\rm dm}$ and
$F_{\star}$ are both included is plotted in panel (b), but now for
a GC starting at a distance $1.5 \hat{r}_{\ast}$ in order to explore a
larger dynamical range in radius. 
We can see that in this case the tendency of sedimentation of the clusters
is significantly suppressed as compared to the one-component estimate
(Eq.~2). In fact, a massive object of mass
$2\times 10^{5}$ M$_{\odot}$ initially at $1.5\hat{r}_{\ast}$ takes
$10$ Gyr to reach the stellar core radius and more than $25$ Gyr
to decay its radius to $1/5$th of its initial value. If the same object is
initially located at the stellar core radius, it would sink
to a radius $0.2 \hat{r}_{\ast}$
in $20$ Gyr approximately. Roughly speaking, for the mass model
under consideration (${\mathcal{R}}=7$,
$\beta=2.5$ and $\hat{v}_{c}=20$ km s$^{-1}$), a GC embedded in
the stellar core will reduce its orbital radius by only 
a factor $2$ during its lifetime.
With these parameters the problem
of the dramatic decay of the orbit is overcome.

For comparison and to show the sensitivity of the evolution to changes
on $\beta$, let us compare the orbital evolution between a cluster 
in a model ${\mathcal{R}}=7$,
$\beta=2.5$ and $\hat{v}_{c}=15$ km s$^{-1}$ (dashed-line in 
Fig.~\ref{fig:evolution}b), 
and in a model ${\mathcal{R}}=7$,
$\beta=1.7$ and $\hat{v}_{c}=20$ km s$^{-1}$ (Fig.~\ref{fig:evolution}c).
For the first model,
a cluster initially at the
stellar core radius will take $7.7$ Gyr to shrink its radius a 
factor $2$, whereas in the second one, this requires only $5$ Gyr.
For the latter model a noticeable radial contraction of the population
of GCs as a whole is expected, which is not observed in Fornax.

The following sets of parameters satisfy the condition that 
all the clusters that
are initially at a spatial radius $>1.5 \hat{r}_{\star}$ from 
the centre of the host dwarf galaxy,
remain outside a sphere of radius $0.75 \hat{r}_{\star}$ after $10$ Gyr:
for a core radius $\hat{r}_{\rm dm}=1.0$ kpc, we need ${\mathcal{R}}=15$ and 
$\hat{v}_{c}=25$ km s$^{-1}$, whereas for $\hat{r}_{\rm dm}=1.5$ kpc,  
a value ${\mathcal{R}}=7$ and velocity $\hat{v}_{c}=15$ km s$^{-1}$ suffice.
The simplest hypothesis for the GC population is that they started
out with the same spatial distribution as the underlying stellar population.
However, the initial spatial distribution of GCs is unknown and could have
been more diffuse than at present. Still, even for a GC population
initially distributed as the stars, GCs could have avoided sedimentation toward the
core of Fornax for a range of halo parameters, as noted above. 
Of course, the evolution of the radial distribution of
a typical dSph or dE globular cluster system requires Monte Carlo
simulations in order to include the mass function of the globular
clusters which could depend in principle on the radius (see Lotz et al.~2001).

It is worthwhile to emphasize here that the mass models with 
$\hat{v}_{c}\sim 20$ km s$^{-1}$ are consistent with the observed
stellar kinematics, but values $\hat{v}_{c}\sim 25$ km s$^{-1}$ are
marginally inconsistent. In Fig.~\ref{fig:circularvel}, 
the circular velocity of the halo for the mass model
${\mathcal{R}}=7$, $\beta=2.5$ and $\hat{v}_{c}=15$ km s$^{-1}$, is
shown. The total mass-to-luminosity ratio interior to $2$ kpc
is $20$ for this model. A comparison with the models of Lokas (2002)
and Walker et al.~(2005)
reveals that this model is fully compatible with the observed
kinematics of the stars (see also Kazantzidis et al.~2004).
In fact, Walker et al.~(2005) found that two-component King models
can reproduce the radial velocity of the new sample of stars belonging
to Fornax. For these models, the
cumulative mass-to-light ratio within the core radius is $7$--$10$
(see their Fig.~9).
Since we are using two-component models similar to those used by
Walker et al.~(2005), we do not repeat the calculations of the 
expected velocity dispersion profile here.

In conclusion, the preservation of the clusters is assured if Fornax
has a cored dark halo with physically reasonable King core radii of
$\hat{r}_{\rm dm}\sim 1.5$ kpc and ${\mathcal{R}}\gtrsim 7$.
This mass model implies a central velocity dispersion for dark matter
particles of $\sim 30$ km s$^{-1}$ and a total mass $\sim 1.2\times 10^{9}$ 
M$_{\odot}$.
These parameters are also plausible for other dSph and dE 
(see \S \ref{sec:discussion}). Notice that a very general result of galactic 
formation scenarios is that typical
extents of the dissipationless dark matter halo are expected to be larger 
than those of the baryonic component,
which collapse into the bottom of the potential well by dissipative
processes, by factors of $\gtrsim 10$
(e.g. Fall  \& Efstathiou 1980; Kregel et al. 2005).

\begin{figure*}
  \includegraphics[width=0.8\textwidth]{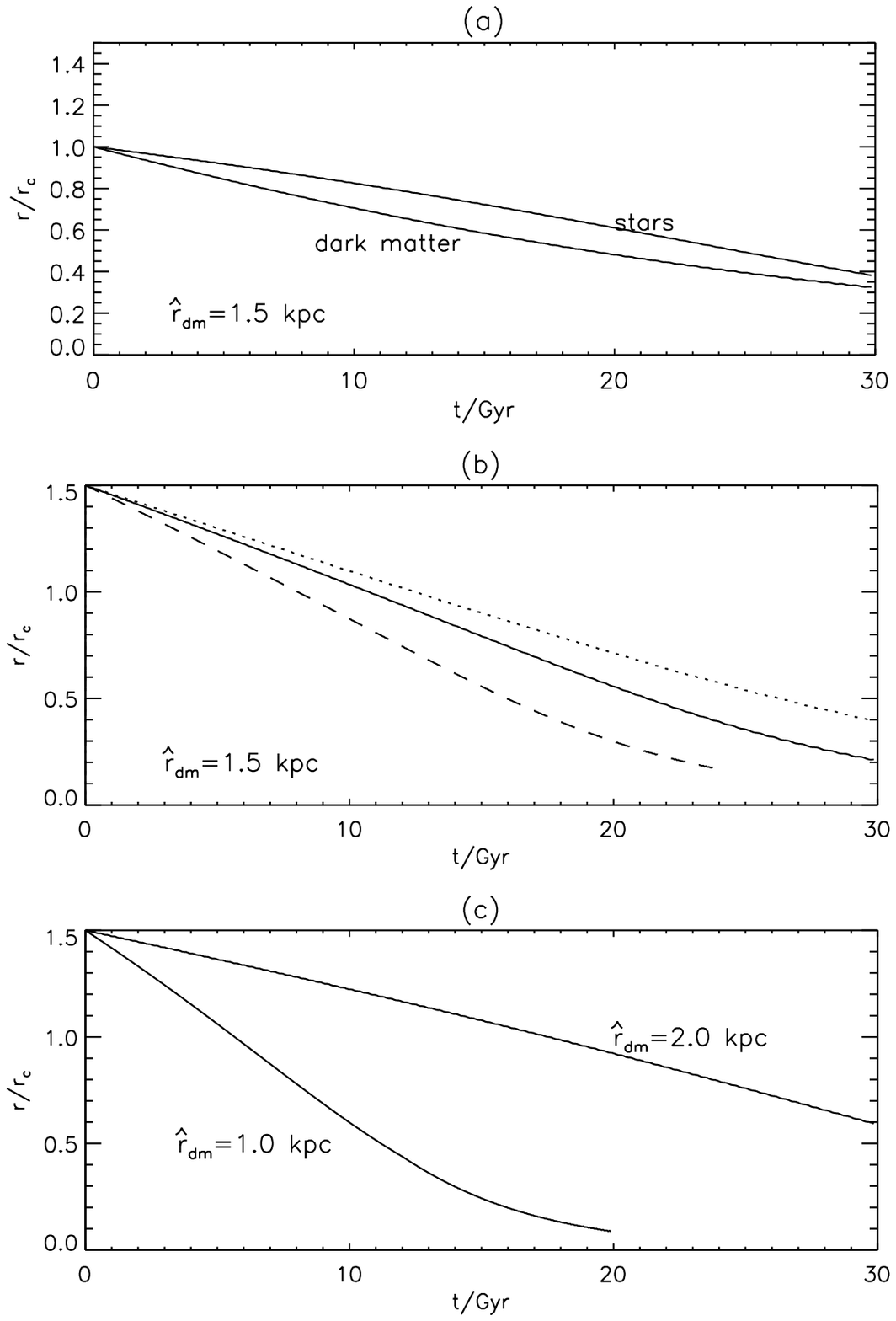}
  \caption{Evolution of the radius of the globular cluster
 for different parameters of the model. Orbital decay if
only stars or dark matter contribute to dynamical friction
for  ${\mathcal{R}}=7$
and $\hat{v}_{c}=20$ km s$^{-1}$ (panel a). 
Decay including both the stellar and dark matter components (panel b),
for $({\mathcal{R}},\hat{v}_{c})=(15,20),(7,20),(7,15)$,
from top to bottom. 
In panel (c) the dependence on the dark matter core radius is shown
for $({\mathcal{R}},\hat{v}_{c})=(7,20)$ .}
  \label{fig:evolution}
\end{figure*}

\subsection{Other profiles for the dark halo: Distinguishing cores from
cuspy haloes}
\label{sec:profiles}
In the previous section we have demonstrated that the solution
to the GC decay problem may reside in a halo with a core 
somewhat larger than that of the stellar population.
It is therefore natural to check whether a cored halo is a 
necessary condition or if cuspy haloes have the same interesting potential.
In this section we will show that clusters cannot avoid sinking
into the central region of Fornax if the dark halo is cuspy.
Therefore, in the lack of any viable mechanism
responsible for the dynamical
heating of the GC system, the requirement of their survival
may serve as a useful tool to discriminate between cuspy or cored haloes.
In \S \ref{sec:discussion} 
other observational constraints regarding the 
dark matter haloes of dSph are given.
Before doing so, it is convenient to consider
the DF timescale for GCs embedded in dark haloes with density profiles
different from the King spheres considered so far. When comparing 
dark matter profiles with different functional forms, we shall do so 
at fixed observational consequences,
i.e., requiring final galactic models having similar velocity dispersion 
profiles for the stars. We will see that cored haloes are favored 
against cuspy haloes in dSphs.

We use a broad family of density profiles for the
halo:
\begin{equation}
\rho(r)=\frac{\rho_{s}}{(r/r_{s})^{\alpha}[1+(r/r_{s})^{\gamma}]
^{\beta}}.
\end{equation}
The parameters ($\alpha,\beta,\gamma)$ determine the shape 
of the density profile. 
The mass density distribution in the inner parts
is described by a power-law $\rho\sim r^{-\alpha}$. 
The set of parameters $(0,1,2)$ corresponds to the pseudo-isothermal profile.
Our aim here is to find out the friction timescale
for cuspy haloes, say $\alpha> 0$.
For this purpose we consider two broadly used profiles:
the singular isothermal sphere $(2,0,\gamma)$, 
and the NFW profile $(1,2,1)$, suggested by cosmological N-body 
simulations, e.g.~Navarro et al. (1996).
The profile of the singular isothermal sphere can be also reproduced 
adopting $(0,1,2)$ in the limit $r_{s}\rightarrow 0$
and $\rho_{s}r_{s}\rightarrow$ constant. That is, the singular
sphere can be recovered as the limit of a pseudo-isothermal sphere 
with a vanishing core radius. As we have seen in \S \ref{sec:numerical} 
that a core with a minimum size is needed to explain the present 
configuration of GCs, 
we can anticipate that the singular isothermal sphere will predict an 
excessively short dynamical-friction timescale.

\begin{figure*}
  \includegraphics[width=0.6\textwidth]{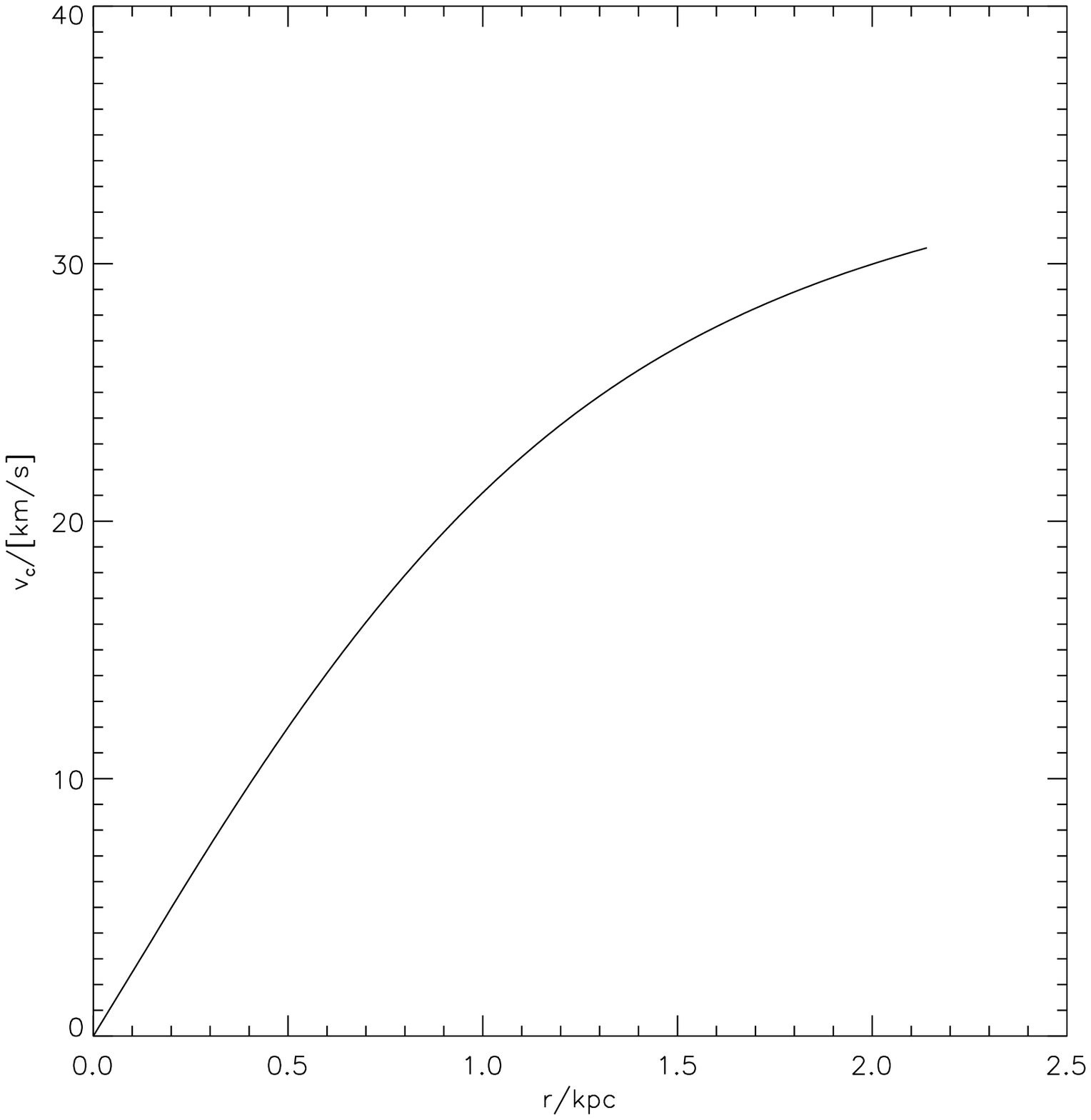}
  \caption{Circular velocity for the dark halo
in the mass model $\hat{r}_{\rm dm}=1.5$ kpc, ${\mathcal{R}}=7$
and $\hat{v}_{c}=15$ km s$^{-1}$.}
  \label{fig:circularvel}
\end{figure*}

In the singular isothermal halo with density
$\rho(r)=v_{c}^{2}/4\pi Gr^{2}$, where $v_{c}$ is the circular velocity,
the time for the orbit of a globular cluster 
to decay from an initial radius $r_{i}$ to the centre is
\begin{equation}
t_{\rm df}=\frac{2.64\times 10^{2}}{\ln\Lambda}
\left(\frac{r_{i}}{2\, {\rm kpc}}\right)^{2}
\left(\frac{v_{c}}{250\, {\rm km\,s}^{-1}}\right)
\left(\frac{10^{6}\,{\rm M}_{\odot}}{M_{p}}\right) \, {\rm Gyr},
\label{eq:tauiso}
\end{equation} 
(e.g., Binney \& Tremaine 1987). We see that, 
in contrast to the cored halo case,
where the orbital radius decay exponentially in time,
in the singular isothermal sphere, as well as in a NFW halo 
(Taffoni et al. 2003),
the body experiencing dynamical friction reaches the centre in a finite 
time. This must be borne in mind when comparing the timescales in 
this section to what was obtained for King spheres.
Note, however, that the difference between $t_{\rm df}$ and $\tau_{5th}$
is not quantitative important in these models because the DF
timescale is determined mainly by the initial orbital radius where the density
is smallest. In fact, $\tau_{5th}=0.96t_{\rm df}$ in the singular isothermal
sphere.

The previous formula has proved very
useful in the study of decay timescales of massive objects in the outer
parts of the halo where the main contribution to the dynamical friction
comes form dark matter particles following a $\rho\sim r^{-2}$ profile.
Let us evaluate Eq.~(\ref{eq:tauiso}) assuming that the halo of Fornax
follows the singular isothermal sphere at any radius. We need to estimate
the associated circular velocity $v_{c}$ for this halo.
From the mass models of Lokas (2002) with varying index $\alpha$
and stellar anisotropy, 
it turns out that a singular
isothermal halo with a value $v_{c}$ larger than $30$ km s$^{-1}$
would be in conflict with the observed kinematics of the stars 
in Fornax, since it overestimates the stellar velocity
dispersion. Hence, if generously adopting it as a characteristic 
value of $v_{c}$, 
Equation (\ref{eq:tauiso}) with $\ln\Lambda=3$ predicts 
an uncomfortably short timescale $t_{\rm df}\lesssim 4.5$ Gyr for a GC 
of mass $2\times 10^{5}$ M$_{\odot}$ and initially 
at a distance $r_{i}=0.6$ kpc, to reach the
centre caused by the dynamical friction with the dark matter particles
alone. Only those GCs
$r_{i}>1.5$ kpc are expected to avoid dramatic sinking towards the centre in 
one Hubble time.  In order for GCs placed within the core of
the dSph to survive for $10$ Gyr, one should invoke extremely large 
values  $v_{c}>65$ km s$^{-1}$, which is inconsistent with the data.

The NFW density profile, proposed as a
universal fitting formula for CDM haloes in the hierarchical clustering
scenario (Navarro et al.~1996, 1997), 
is less cusped than the singular isothermal sphere, and
has the form:
\begin{equation}
\rho(r)=\frac{\rho_{s}r_{s}^{3}}{r(r+r_{s})^{2}},
\end{equation}
where the parameters $\rho_{s}$ and $r_{s}$ are the density scale and 
scale radius, respectively. 
Taffoni et al.~(2003) provided an expression for $t_{\rm df}$, assuming
a NFW profile for the dark halo: 
\begin{equation}
t_{\rm df}\simeq 2 \frac{V_{200}r_{i}^{2}}{GM_{p}\ln\Lambda},
\label{eq:taffoni}
\end{equation}
where $r_{i}$ is the initial radius and $V_{200}$ is the circular velocity
at $r_{200}$, which encompasses a mean overdensity of $200$ times
the critical density.
By modeling the moments of the line-of-light velocity distribution
in Fornax, Lokas (2001) obtained a best-fitting virial
mass in the range $M_{200}=1.3$--$1.5\times 10^{9}$ M$_{\odot}$,
depending on the degree of anisotropy assumed for the stellar orbits,
for a concentration parameter $c\approx 20$, as suggested in cosmological
simulations. The specification of the halo mass and concentration 
allows all other parameters to be deduced. In particular,
we get $V_{200}\simeq 17.5$ km s$^{-1}$, and hence, 
Eq.~(\ref{eq:taffoni}) predicts $t_{\rm df}\approx
4.5$ Gyr for a GC of mass $2\times 10^{5}$ M$_{\odot}$
and $r_{i}=0.6$ kpc 
(we continue using $\ln\Lambda=3$).

Bearing in mind that the observational
determinations of core radii and velocity dispersion profiles 
of dSph galaxies are substantially larger than the few percent 
error incurred by the use of the King approximations,
we see that their use does not affect our conclusions. 
In essence, the result is that changing the assumed dark matter 
profile from cored to cusped, within restrictions
imposed by stellar kinematics, the DF timescales change 
by a factor of $\gtrsim 3$, 
allowing one to exclude the latter option,
in spite of the approximations of the analysis or the 
uncertainties in the data.

\subsection{Other galaxies}
Fornax is not the only nearby dwarf galaxy with a population of GCs.
Five GCs were identified in the dwarf elliptical satellite of
Andromeda, NGC 185 (Hodge 1974). Since the stellar core radius of NGC 185 is
only $0.1$ kpc, Tremaine (1976) derived a decay rate four times faster 
than in Fornax. The explanation of an extended dark halo for the
problem of the survival of those GCs may be problematic
as the value of the mass-to-light
ratio for this galaxy of $\sim 3$ is rather low (Held et al.~1992)
and, hence, the interaction with the stars rather than with the dark matter
will dominate the drag, with no much room left to dilute the effect 
of the friction (see the two-component formula, Eq.~\ref{eq:TwoCompDF}).
However, the closest GC to the centre of this galaxy
is at a projected distance of $\sim 6\hat{r}_{\star}\sim 0.6$ kpc, 
well outside the core of
the galaxy. NGC 185 is a clear case where the globular cluster system
is much more extended than the stellar population.
At such distances from the centre, the stellar density drops by a large factor
making dynamical friction inefficient.

Lotz et al.~(2001) examined a sample of dE galaxies for evidence of
dynamical friction in their globular cluster systems and nuclear
properties. They concluded that for the fainter galaxies, some mechanism
should be working against the orbital decay of GCs: 
mass loss via supernova-driven winds, large dark matter cores
greater than $\sim 2$ kpc, the formation
of new star clusters, or tidal stripping.
Our calculations give further support to the suggestion
that dark matter King core radii greater than $\sim 2$ kpc for the fainter
dE galaxies can explain the observed trend.

\subsection{Cores in dwarf galactic haloes}
\label{sec:discussion}

In the previous sections, the study of dynamical friction timescales
in dwarf galactic systems has lead us to conclusions
supporting the idea of constant density cores for the dark matter
haloes of such galaxies. In the present subsection we discuss alternative 
and independent lines of evidence pointing in the same direction.

Observations of rotation curves from the Milky Way
to nearby dwarf galaxies and low
surface brightness galaxies generally show that their dark matter density
profiles have a flat inner core (e.g., Binney \& Evans 2001;
van den Bosch et al.~2000; de Blok \& Bosma
2002; de Blok 2005). This appears to be in disagreement with the results of
numerical simulations of CDM haloes. 
Despite the ongoing debate about the correct interpretation
of the gas rotation curves of low-mass disc galaxies 
(Hayashi et al.~2004; Spekkens et al. 2005), 
cored profiles appear to be favoured over cuspy halo models (e.g., Gentile
et al.~2005). The gas-free
Local Group dSph galaxies provide an independent test for dark matter
models as they are often completely dark matter dominated at all
radii, and it is therefore possible to measure their dark matter content
by treating their baryonic content as a massless tracer population.
If dSph obey the same scaling relation $\rho_{0,{\rm dm}}$ vs 
$\hat{r}_{\rm dm}$ found empirically by
Fuchs \& Mielke (2004) in low surface brightness galaxies, 
a core radius of $\sim 3$ kpc should be expected for a central
density as that inferred for Fornax-like dSph, $\sim 0.02$ M$_{\odot}$ 
pc$^{-3}$.

A great effort has been made in order to constrain the total $M/L$
in dSph. The density profile of the haloes and their
parameters still remain very uncertain, mainly due to
large uncertainties in both the velocity dispersion measurements and
the anisotropy of the stellar velocity dispersion. 
Lokas (2002, 2003) studied a set of dark matter profiles with
different slopes in the inner parts and 
showed that all dark matter profiles yield good fits but only profiles 
with cores are consistent with isotropic orbits.
A {\it King core radius} of $\sim 1.5$ kpc required to 
preserve the GC configuration
in Fornax, as inferred in \S \ref{sec:numerical}, 
can fit reasonably well the stellar
velocity dispersion profile observed in Fornax (e.g., Lokas 2002; 
Stoehr et al.~2002; Kazantzidis et al.~2004; Walker et al.~2005).
Throughout this paper, by core radius $\hat{r}$ we always mean 
the radius at which the projected density falls to half of 
its central value. In cosmological
simulations, it is usual to define the core radius $r_{c,\gamma}$
as the point where the density profile becomes steeper than $\gamma$.
In order to avoid confusion between the various scale radii, 
the following relations are useful
when comparing with cosmological models of structure formation:
$r_{c,-1}\approx 0.75\hat{r}$ and
$r_{c,-0.1}\approx 0.2\hat{r}$. 

There exist other independent lines of evidence for cores in dSph.
The recent identification of a kinetically cold stellar substructure
in the Ursa Minor dSph (Kleyna et al.~2003) strongly suggests that
the dark halo of that galaxy has a central core. Using N-body
simulations Kleyna et al. (2003) showed that this cold structure
would survive for less than $1$ Gyr if the dark halo is cusped.
Only if the dark halo has a uniform density core can the cold 
substructure survive. Additionally, Magorrian (2003) found  
an inner slope $\alpha=0.55\pm 0.35$ for the Draco dSph.

If it is confirmed that dSph galaxies contain cored haloes, then the problem
of the formation of cores becomes universal in dwarfs, implying
that cores are not exclusive of disc galaxies where any
conflict between predictions and observations can be attributed to 
our poor understanding of the complicated physics of baryons, such
as feedback, pressure support or transport of angular momentum by bars. 
This fact would have serious implications for models aimed to explain cores. 
dSphs with extended dark haloes also arise in $\Lambda$CDM simulations
(Stoehr et al.~2002), the question to ask is how these extended
haloes transform their central cusps into cores.

Although the formation of cores is a delicate issue, 
some possible ways have been explored in the literature.
S\'anchez-Salcedo (2003)
proposed that decaying dark matter could explain the formation of
cores in LSB and dSph. Read \& Gilmore (2005) have suggested that 
the sudden impulsive loss of baryonic mass in a dSph 
could produce a redistribution of 
matter, transforming a central density cusp into a near-constant
density core. Jin et al.~(2005) show that a halo which is composed
of massive black holes and which initially has a NFW density profile,
can be transformed into a cored halo through dynamical evolution. 
The core radius of these haloes is $\sim 1$ kpc and resemble Draco's halo. 

The present study puts constraints on the epoch of core formation. In fact,
if cored haloes are solely responsible for the
survival of the present-day GC configuration, as suggested in this paper,
the formation of the core should have proceeded 
in an early phase of the galaxy's formation, at least $6$ Gyr ago,
to prevent excessive drag of Fornax GCs.

\section{The alternative: MOND}
\subsection{MOND in dSph and dynamical friction}
\label{sec:introMOND}
An interesting alternative theory to the existence of massive
dark matter haloes is a modification of the standard Newtonian gravity
for accelerations below some characteristic value, $a_{0}=1$--$2\times
10^{-8}$ cm s$^{-1}$ (Milgrom 1983). The success of MOND in
reproducing the observed velocity rotation curves of spiral galaxies without
dark matter haloes is amazing, only in about $10\%$ of the roughly $100$
galaxies considered in the context of MOND does the predicted rotation
curve differ significantly from that observed 
(e.g., Sanders \& McGaugh 2002).
Although there are some aspects 
where MOND predictions are not fully satisfactory in
disc galaxies (e.g., S\'anchez-Salcedo
\& Hidalgo-G\'amez 1999; Blais-Ouellette et al.~2001; 
Binney 2003; Gentile et al. 2005; 
S\'anchez-Salcedo \& Lora 2005), MOND remains an intriguing alternative 
to dark matter at galactic scales. In principle, the parameter $a_{0}$ should
be universal and, having determined its magnitude, one is not allowed
to leave it as a free parameter. However, the derived value depends upon
the assumed distance scale. Here, the MOND acceleration parameter is
assumed to be the value derived in Begeman et al.~(1991), rescaled to
the new distance scale, i.e.~$a_{0}=0.9\times 10^{-8}$ cm s$^{-2}$
(Bottema et al.~2002). 

Because of their large sizes, dSph galaxies also lie in the regime
of small accelerations and hence, provide a laboratory to test MOND.
In the previous section we show that the survival of GCs in dSph
may provide new insight into the halo properties. We will consider now
the implications for MOND. Ciotti \& Binney (2004) have already pointed out
that the existence of GCs in dwarfs may be a serious difficulty for
MOND. In order to address the severity of this potential problem, 
we wish to compare the timescale of GC orbital decay
in MOND to that in the dark matter scenario in the case study of Fornax, a dSph galaxy
with well-determined photometric (structural) parameters, as a
diagnostic of MOND. 

Lokas (2002) has applied MOND to Fornax, Draco and Ursa Minor dwarfs.
She found that if the stellar-to-mass ratio is fixed to $1$ 
M$_{\odot}/$L$_{\odot}$, the best fitting values of $a_{0}$, 
which is a universal
constant in MOND, lie in the range acceptable to explain the rotation
curves of spirals, but much higher in the case of Draco and Ursa Minor.
Conversely, adopting a value $a_{0}=1.2\times 10^{-8}$ cm s$^{-2}$, the
best fitting model for Draco corresponds to a $M/L=32\pm 10$ 
M$_{\odot}/$L$_{\odot}$. Therefore, Draco needs a dark component even in MOND
(see also Lokas, Mamon \& Prada 2005), weakening the reliability
of the MOND paradigm, but this fact by itself does not rule out 
MOND\footnote{Note that there is consensus in
the discrepancy between the baryonic mass and
the dynamical mass in clusters of galaxies in MOND.}.
In Fornax, the best-fitting model requires a reasonable stellar-to-mass
ratio of $1.8\pm 0.4$ M$_{\odot}/$L$_{\odot}$ and no dark matter at all.
Since we are primarily interested in the GCs evolution in the context
of Fornax, we will adopt in our study the stellar-to mass ratio 
within this range.

In isolated spherical systems, the nonlinear MOND field 
equation for the gravitational potential reduces to
an algebraic relation between the real acceleration $\bmath{g}$ 
and the Newtonian
acceleration $\bmath{g}_{N}$, $\mu\left(|\bmath{g}|/a_{0}\right) 
\bmath{g}=\bmath{g}_{N}$ 
(Brada \& Milgrom 1995), where 
$\mu(x)$ is a smooth function which is not specified, but approaches $1$
in the limit of large $x$ and approaches $x$ in the limit of small
$x$ (the deep MOND regime). 
It can be seen that the acceleration $g=v_{c}^{2}/r$ felt by a star 
in the core of Fornax 
is significantly smaller than $a_{0}$ and thus, for the dynamics of an
object in Fornax the deep MOND regime would apply.  
Because of MOND's nonlinearity, a system's internal dynamics can be altered
by an external field of acceleration $g_{\rm ext}$ within which it 
is immersed (Bekenstein \& Milgrom 1984).
A measure of the external field effect in a dwarf at a position
$D$ from the parent galaxy is the parameter $\eta\equiv 1.5 
(\sigma_{\ast}/V)^2(D/\hat{r}_{\star})$ (Milgrom 1995),
with $V$ the galactic rotational velocity at $D$, which 
coincides with the asymptotic rotation velocity $V_{\infty}$ for
all the dwarfs.
In the quasi-Newtonian limit, i.e.~when $\eta\ll 1$ and
all accelerations relevant to the dwarf dynamics are smaller than $a_{0}$,
the dynamics becomes Newtonian but with
a larger effective gravitational constant of $Ga_{0}/g_{\rm ext}$.
Hence, we need to know the degree of isolation of the dwarf to
investigate the dynamics of GCs. Note also that
given the luminosity profile and the stellar $M/L$,
the circular velocity curve may be different for the isolated case
and the quasi-Newtonian case.

By considering general principles of statistical mechanics,
Ciotti \& Binney (2004) found that the dynamical-friction time in an
isolated system in the weak acceleration limit,
is shorter by a factor $a_{0}^{2}/(\sqrt{2}g^{2})$
over the value it would have in a Newtonian system with the same
stellar mass and a fixed auxiliary gravitational potential.  
The reason for this is that encounters at impact parameters
comparable to the half-mass radius are dominant and provide the major
contribution to the DF in MOND.
The approach developed by Ciotti \& Binney can be extended 
easily to the quasi-Newtonian limit if the calculation is carried over
with the substitution $g\rightarrow g_{\rm ext}$ and hence,
the DF time in the quasi-Newtonian limit is shorter than in the 
Newtonian case by a factor $a_{0}^2/(\sqrt{2}g_{\rm ext}^{2})$. 

In Fornax $\eta=0.85$ 
and therefore it is a borderline case as regards MOND isolation.
Neither limit is valid in this case but the two limits should give
similar values for the friction. In the next section we will explore
two extreme situations: Fornax in isolation and Fornax in the
quasi-Newtonian limit. Since the force law depends on the degree
of isolation, it is important to keep in mind that the stellar
mass-to-light ratio required to fit the observed velocity dispersion
profile is not the same in both limits. Hence, once given the stellar
kinematics and the luminosity profile, the mass density is different
in the isolated weak-field limit and in the quasi-Newtonian regime. 
Hereafter, where appropriate, the scripts ISO will denote the isolated
regime and QN the quasi-Newtonian limit.

In the derivation of the friction force acting on a GC at
a distance from the centre of the dwarf $r$, we will adopt the local
approximation. That is, we extend the results of Ciotti \& Binney
(2004) evaluating the force with the local variables. The local
approximation assumes that there are no strong gradients in the
properties of the stellar background. However, it is liable to 
objection because of the 
long-range behaviour of the MOND law. In Newtonian dynamics much
effort has been done to prove that the local approximation is a
realistic assumption for most astrophysical scenarios. 
In the context of MOND this remains to be
checked. Therefore, although our results should be taken
with caution, the DF force based on the local approximation might
have uncertainties of as much as a factor $2$ in a cored dSph.

\subsection{MOND Orbital-decay timescale in galaxy cores - Isolated galaxy under deep MOND regime} 

As said in the previous subsection, in the framework of deep-MOND regime,
the DF force exerted by the stars on a massive object 
orbiting within the bulk of an isolated dwarf 
is enhanced by a factor
$a_{0}^{2}/(\sqrt{2}g^{2})= a_{0}^{2}r^{2}/(\sqrt{2}v_{c}^{4})$
over its Newtonian value.
Hence, the DF force at a radius $r$ from the centre of the dwarf
in the local approximation is:
\begin{equation}
F_{\star}^{\rm ISO}(r)=-\frac{4\pi \ln \Lambda G^{2} \rho_{\star}(r) M_{p}^{2}}
{v_{c}^{2}} \left(\frac{r^{2}a_{0}^{2}}{\sqrt{2}v_{c}^{4}}\right)
\left[{\rm erf}\left(\frac{v_{c}}{\sqrt{2}\sigma_{\ast}}
\right)-\sqrt{\frac{2}{\pi}}\frac{v_{c}}{\sigma_{\ast}}\exp\left(
-\frac{v_{c}^{2}}{2\sigma_{\ast}^{2}}\right)\right].
\label{eq:isolationMOND}
\end{equation}
In the deep MOND limit, the circular velocity is given
by $v_{c}(r)=(Ga_{0}M_{\star}(r))^{1/4}$, and the mean one-dimensional
velocity dispersion by $\sigma_{\star}^{4}=(4/81) Ga_{0}M_{T}$ 
(Bekenstein \& Milgrom 1984), where $M_{T}$ is the total (baryonic) mass.
Substituting the expression for $v_{c}(r)$, with
$M_{\star}(r)\simeq 4\pi\rho_{0,\star}r^{3}/3$, 
into Eq.~(\ref{eq:isolationMOND}), 
we find that massive objects in the inner core, 
i.e.~at those radii such as $M_{\star}\ll M_{T}$,
spiral towards the dwarf centre at a constant rate according to:
\begin{equation}
\frac{dr}{dt}=-0.45 \frac{G a_{0}M_{p}\ln\Lambda}{\sigma_{\star}^{3}},
\end{equation}
where local circularity of the orbit has been assumed.
The dynamical-friction timescale for MOND of a body initially at
circular orbit with radius $r_{i}$ is then
\begin{equation}
\tau_{5th}^{\rm ISO}=1.8\frac{r_{i}\sigma_{\star}^{3}}{Ga_{0}M_{p}\ln\Lambda}.
\label{eq:tauISO}
\end{equation}
Starting at a radius inside the core, say $0.3$ kpc,
this DF timescale is extremely short, $\sim 0.09$ Gyr, for our fiducial values
$\sigma_{\star}=11$ km s$^{-1}$, $M_{p}=2\times 10^{5}$
M$_{\odot}$ and $a_{0}=0.9 \times 10^{-8}$ cm s$^{-2}$. As noticed first by
Ciotti \& Binney (2004), the DF timescale is of the order of the dynamical
crossing time. In fact, the orbital evolution is so fast that our 
assumption that 
the decay is through circular orbits is not fully justified. In the next 
section, however,
we solve numerically the orbit of a massive GC in Fornax, without
demanding local circularity.
For a given $\sigma_{\ast}$ taken from observations of a certain dSph,
since $\tau_{5th}$ is inversely proportional to $a_{0}$, our adopted
scaled value of $a_{0}$, which is smaller than others previously
considered in the literature (e.g., The \& White 1987), 
gives a generous estimate of the DF timescale.

\subsection{Dwarf galaxy in an external field in the quasi-Newtonian limit}
\label{sec:qn}
In the quasi-Newtonian limit, the form of the force is similar to the
Newtonian case but replacing $G\rightarrow Ga_{0}/g_{\rm ext}$, plus
an extra-factor of $\sqrt{2}$ (see \S \ref{sec:introMOND}): 
\begin{equation}
F_{\star}^{\rm QN}(r)=-\frac{4\pi \ln \Lambda G^{2} \rho_{\star}(r) M_{p}^{2}}
{v_{c}^{2}} \left(\frac{a_{0}^{2}}{\sqrt{2}g_{\rm ext}^{2}}\right)
\left[{\rm erf}\left(\frac{v_{c}}{\sqrt{2}\sigma_{\ast}}
\right)-\sqrt{\frac{2}{\pi}}\frac{v_{c}}{\sigma_{\ast}}\exp\left(
-\frac{v_{c}^{2}}{2\sigma_{\ast}^{2}}\right)\right].
\label{eq:qnMOND}
\end{equation}
A similar analysis to the Newtonian one 
carries over to get a formula for $\tau_{5th}^{QN}$
for bodies orbiting well inside the core:
\begin{equation}
\tau_{5th}^{QN}=1.88 \left(\frac{g_{\rm ext}}{a_{0}}\right)
\left(\frac{\sigma_{\ast}\hat{r}_{\ast}^{2}}{GM_{p}\ln\Lambda}\right),
\end{equation}
resulting in an expression formally identical to Eq.~(\ref{eq:newtonian}), with
${\mathcal{R}}=0$, except for the factor
$\sqrt{2}(g_{\rm ext}/a_{0})$. For Fornax standard values, 
$\tau_{5th}^{QN}=0.3$ Gyr, i.e.~a factor 
$\sim (V_{\infty}/\sigma_{\star})^{2}(\hat{r}_{\star}/D)(\hat{r}_{\star}/
r_{i})= 1.5\eta^{-1}(\hat{r}_{\star}/ r_{i})=1.7(\hat{r}_{\star}/ r_{i})$ 
larger than in the isolated case.
However, we warn that these timescales were derived for an object in
the inner core of a galaxy and, therefore, it is not clear at this stage
whether this difference between the timescales in the isolated and 
quasi-Newtonian cases also holds for a body placed initially
 outside the core radius.  In fact, let us estimate 
the ratio between the magnitudes of the local DF force 
in the isolated case (Eq.~\ref{eq:isolationMOND}) and
in the quasi-Newtonian case (Eq.~\ref{eq:qnMOND}) 
at a certain radius $r$. For simplicity of the discussion, suppose
that in both cases the circular velocities $v_{c}$ at $r$ are roughly
the same, which implies that the $M/L$ must be rearranged to have similar
stellar kinematics, i.e.~similar $\sigma_{\star}$, in both limits.  
After some manipulations, it is easy to show that the ratio of the
forces is given by:
\begin{equation}
\frac{F_{\star}^{\rm QN}}{F_{\star}^{\rm ISO}}=
\frac{v_{c}^{2}}{rg_{\rm ext}}=\left(\frac{v_{c}}{V_{\infty}}\right)^{2}
\left(\frac{D}{r}\right). 
\label{eq:ratioQI}
\end{equation}
For a typical Fornax-like galaxy, 
the characteristic circular velocity in Fornax
is $25$ km s$^{-1}$ at $1.5\hat{r}_{\star}$. 
Equation (\ref{eq:ratioQI}) implies that 
$F_{\ast}^{\rm QN}$ is a factor $\sim 2$ larger than $F^{\rm ISO}$ 
at $1.5\hat{r}_{\star}$. In fact, we will show in the next section
that the spiraling of a GC initially at $1.5\hat{r}_{\star}$ 
is faster in the quasi-Newtonian limit
than in the isolated weak-field limit, in a Fornax-like galaxy with 
$\eta\sim 1$.
  
\begin{figure*}
  \includegraphics[width=0.6\textwidth]{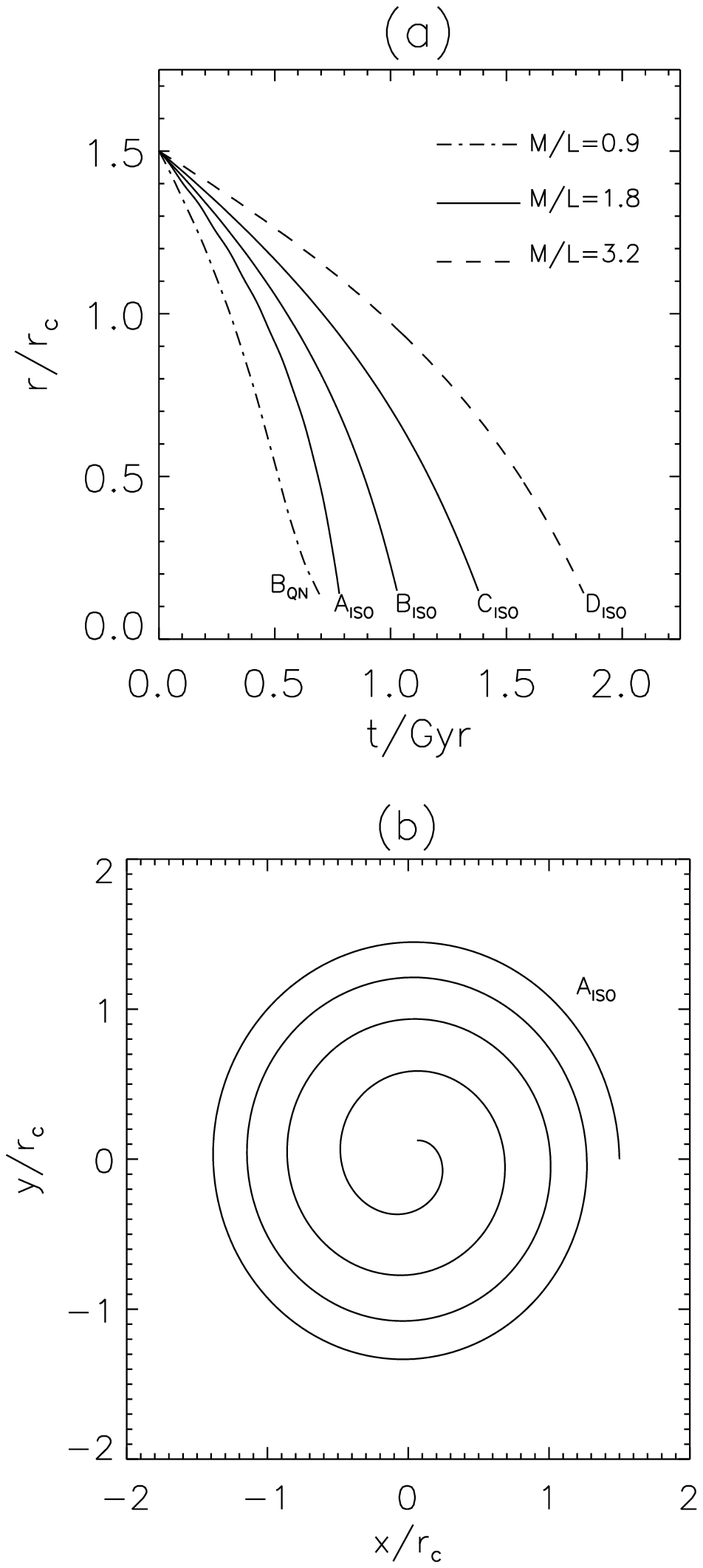}
  \caption{Panel (a): Temporal evolution of the radial position of a 
massive object of $2\times 10^{5}$ M$_{\odot}$
in MOND for different values of the mass-to-light ratio and velocity
dispersions $\sigma_{\ast}=10, 12, 14$ and $15$ km s$^{-1}$ for curves
A, B, C and D, respectively. Panel (b): 
Orbit in the $(x,y)$ plane for case A.}
  \label{fig:MONDevolution}
\end{figure*}

\subsection{Determination of the orbital evolution in MOND}

Using Eq.~(\ref{eq:xavierapprox}) as an approximation for the density profile
of Fornax, and assuming that the stellar orbits are isotropic
with a roughly constant velocity dispersion $\sigma_{\star}$,
the orbital evolution of a GC of mass $2\times 10^{5}$ M$_{\odot}$
has been integrated numerically 
in the local approximation in both regimes discussed above:
isolated weak field limit and quasi-Newtonian limit.
In principle, since in the MOND paradigm there is no dark matter,
the stellar mass-to-light ratio is the only free parameter and
thus, it also fixes the value of $\sigma_{\star}$.
However, in order to explore the sensitivity of the results on the velocity
dispersion, the orbit was calculated taking $\sigma_{\star}$ as an 
independent parameter spanning the reasonable range of $10$--$15$ km s$^{-1}$.

Figure \ref{fig:MONDevolution} shows the in-spiral of the GC 
sinking because of the DF due
to the stars only, for different stellar $M/L$ ratios.
As Eq.~(\ref{eq:tauISO}) suggests, the decay rate in the isolated regime
depends strongly on the adopted value for $\sigma_{\ast}$.
In this regime, the GC reaches $1/5$th 
the initial radius in less than $\sim 4$ orbits.
The GC evolves maintaining its orbit circular until it reaches 
$\sim 0.5 \hat{r}_{\ast}$ (see Fig.~\ref{fig:MONDevolution}b).

A comparison of the curves labeled by $B_{\rm QN}$ and $B_{\rm ISO}$
reveals that the in-spiral takes a longer time in the isolated limit than
in the quasi-Newtonian limit. As already said, the stellar $M/L$ are
different in order to have comparable kinematics (i.e.~similar
rotation curves within $1.5\hat{r}_{\star}$) in both cases.
The time required for
a GC initially at $1.5\hat{r}_{\star}$ to reach the $0.3\hat{r}_{\star}$
is $1.6$ times larger in the isolated weak acceleration limit 
than in the quasi-Newtonian regime. Note however, that the in-spiral
within $0.3\hat{r}_{\star}$ is faster in the isolated case in accordance
with our discussion in \S \ref{sec:qn}. Indeed,
the decay in the QN limit is exponential at small radii 
(see Fig.~\ref{fig:MONDevolution}).

The slowest in-spiral corresponds to the case with the highest 
$\sigma_{\ast}$ and $M/L$ parameters. Notice that a ratio $M/L=3.2$
lies outside $2\sigma$ the value quoted by Lokas (2002) in Fornax
under MOND. But even if adopting these parameters for MOND, 
a massive GC initially in circular orbit at $1.5\hat{r}_{\star}$
reaches the centre of the dwarf in $\sim 2$ Gyr.
Hence, MOND predicts a complete sedimentation of the GCs of Fornax.
In fact, without adopting {\it ad hoc} conditions, MOND is unable
to be compatible with the observed possession of GCs by Fornax.
Consequently, one should invoke external heating effects to rescue MOND.
However, the effects of this heating mechanism should be observable as
distortions in the photometry, for which, as mentioned previously (\S 
\ref{sec:statement}), there is no evidence.

The possession of GC by dwarf galaxies (dE and dSph)
is likely challenging for the MOND theory.
One could ameliorate the problem of the orbital decay just by 
decreasing the universal acceleration $a_{0}$ and 
adding a classical dark halo. But one needs
to reduce $a_{0}$ up to a value $\sim 10^{-9}$ cm s$^{-2}$
for which the classical Newtonian
dynamics at galactic scales is recovered.  
This solution has no astrophysical interest
because then a dark component has to be added as well to explain the
missing mass problem in spiral galaxies. This component will become
the main explanation for the missing mass at galactic scales
and not only at cosmological scales (Pointecouteau \& Silk 2005).

\section{Conclusions}
Long before relevant data was available and using a simple but
transparent scaling of the dynamical friction timescale,
Tremaine (1976) warned that nuclei
should be formed in the centre of dSph galaxies by coalescence of
GCs. He noticed that NGC 185, NGC 147 and Fornax
all have GCs but no nucleus. More recently,
and for a sample of $51$ dwarf elliptical galaxies in nearby clusters,
Lotz et al.~(2001) examined the radial distribution of GCs
and found that for the fainter dE's, the predicted nuclei was several
magnitudes brighter than observed.
Since the composite radial distribution of the GCs
follows the exponential stellar profile of dE's at present day,
the simplest hypothesis is that the GCs initially
followed the exponential profile of the underlying stellar population
and that the dynamical friction force was partly inhibited.
Note that the drag should not be completely suppressed in order
to explain the
significant deficit of bright, massive GCs in the
inner regions of the sample of Lotz et al.~(2001).
In Local Group dSph galaxies, the problem of the survival of
enhanced density substructure with cold kinematics against dynamical
friction has been revived recently (Kleyna et al.~2004; Walker et al.~2006).
For the less massive galaxies, the tidal field could play
a role in heating their GCs but they must be tidally disrupted and
suffer significantly mass loss ($50$--$90\%$) in order to counter
the drag of dynamical friction. If so, these galaxies should present
a rising projected velocity dispersion profile beyond a
critical radius caused by tidal stripping (Read et al.~2006).
In contrast, the Local Group dSph galaxies show a very flat or
falling projected velocity dispersion, which indicates that tidal
stripping is unimportant interior to $\sim 1$ kpc.

In this paper we have shown that all these features can be
explained by considering dark haloes as having density cores.
Taking Fornax as a reference case, we have explored under which
structural parameters of the dark halo is the orbital decay of
GCs minimized (see also Goerdt et al.~2006 for an independent 
and complementary study).
Using self-consistent dynamical models where both the stars and
dark matter contribute to the frictional drag, we have computed
the radial evolution of a massive GC initially on a circular orbit.
We find that the contributions to dynamical friction
of stars and dark matter are both comparable in dwarf galaxies.
The survival, against dynamical friction, of GCs
requires a King core radius for the dark halo 
slightly larger than that of the underlying stellar population.
In the particular case of Fornax, we infer a dark halo with
the following parameters: a King core radius of $\sim 1.5$ kpc
(or $r_{c,-0.1}=0.3$ kpc),
a central velocity dispersion of $\sim 30$ km s$^{-1}$ and a
total mass of $1.2\times 10^{9}$ M$_{\odot}$.
In a dark matter halo with a cuspy inner profile, the dynamical
friction timescale is $\lesssim 4.5$ Gyr. 
The indirect evidence of a central density core in dSph has
important implications for the formation of dSph galaxies and for cosmology,
and gives new insight to the ongoing debate on the inner slope of
the dark halo profiles in low surface brightness galaxies.
The present study puts constraints on the epoch of core formation. 
If cored haloes are solely responsible for the
survival of the present-day GC configuration, 
the formation of the core should have proceeded 
in an early phase of the galaxy's formation, at least $6$ Gyr ago,
to prevent excessive drag of Fornax GCs. 

Under the MOND hypothesis, 
dynamical friction timescales for dwarf galaxies are necessarily short, 
$\sim 1$ Gyr. Therefore, alternative explanations must be found 
for the observed GC systems. No such entirely satisfactory
alternative has been presented to date,
making old GC systems in dwarfs a strong objection to MOND under its
current formulation.

\section{Acknowledgments}
We thank an anonymous referee for a thorough revision of the original 
manuscript resulting in an improved final version and Matt Walker
for providing us the updated profile of the stellar velocity dispersion 
observed in Fornax. The work of X.~Hernandez was partly supported 
by DGAPA-UNAM grant No IN117803-3 and CONACYT grants 42809/A-1 and 42748.

\end{document}